\documentclass[12pt]{article}
\usepackage{amsmath}
\usepackage{graphicx}
\usepackage{color}
\begin{document}
\baselineskip=18 pt
\begin{center}
{\large{\bf Comment on ``Ground state of a bosonic massive charged particle in the presence of external fields in a G\"{o}del-type space-time [Eur. Phys. J. Plus (2018) {\bf 133} : 530]''}}
\end{center}

\vspace{.5cm}

\begin{center}
{\bf Faizuddin Ahmed}\footnote{faizuddinahmed15@gmail.com ; faiz4U.enter@rediffmail.com}\\
{\bf Ajmal College of Arts and Science, Dhubri-783324, Assam, India}
\end{center}

\vspace{.5cm}

\begin{abstract}

We point out incorrect equations derived in a paper published in this journal Ref. \cite{Silva} (E. O. Silva, Eur. Phys. J. Plus (2018) {\bf 133} : 530) for the Klein-Gordon equation with both the Aharonov-Bohm and Coulomb-type potentials in a G\"{o}del-type space-time. We derive the final form of the radial wave equation of Klein-Gordon equation in the Som-Raychaudhuri space-time with these potentials and show that the derive equation here jeopardize the equation obtained in Ref. \cite{Silva} and therefore, the energy eigenvalues presented there is incorrect.

\end{abstract}

{\it Keywords}: G\"{o}del-type space-time, Klein-Gordon equation, electromagnetic potential, energy spectrum, wave-functions.

{\it PACS Number:} 04.20.Cv, 03.65.Pm, 03.65.Ge

\section{Introduction}

In a recent paper in this journal, E. O. Silva \cite{Silva} have studied the relativistic quantum dynamics of a spinless bosonic massive charged particle interacting with both the Aharonov-Bohm and Coulomb-type potentials in the background of a G\"{o}del-type space-time. The author has obtained the final form of the radial wave equation of the Klein-Gordon equation in the Som-Raychaudhuri space-time with these potentials, and evaluated the energy eigenvalues and corresponding normalized eigenfunctions. We show that the derive equations Eq. (4) and Eq. (6) in Ref. \cite{Silva} are incorrect, and two additional terms are missing in Eq. (6). Therefore, the further analysis of the drived equation as well as the energy eigenvalues in Ref. \cite{Silva} are incorrect. In this paper, taking into account all the physical parameters considered in Ref. \cite{Silva}, we derive the final form of Klein-Gordon equation in the Som-Raychaudhuri space-time with the above mentioned potentials as follow.

The Som-Raychaudhuri space-time is described by the following line element (with $c=\hbar=1$):
\begin{equation}
ds^2=-(dt +\Omega\,r^2\,d\phi)^2+r^2\,d\phi^2+dr^2+dz^2,
\label{1}
\end{equation}
where $\Omega$ characterizes the vorticity parameter of the space-time.

The relativistic quantum dynamics of spinless free-particle of mass $M$ is described by the Klein-Gordon equation:
\begin{equation}
\frac{1}{\sqrt{-g}}\,\partial_{\mu} [\sqrt{-g}\,g^{\mu\nu}\,\partial_{\nu}]\,\Psi=M^2\,\Psi,
\label{2}
\end{equation}
with $g$ is the determinant of metric tensor with $g^{\mu\nu}$ its inverse, and $\partial_{\mu}$ is the ordinary derivative.

For the space-time (\ref{1}), the KG-equation (\ref{2}) becomes \cite{JC,ZW}
\begin{equation}
[-\partial_{t}^2+\frac{1}{r}\,\partial_{r}\,(r\,\partial_{r})+(\frac{1}{r}\,\partial_{\phi}-\Omega\,r\,\partial_{t})^2+\partial_{z}^2-M^2]\,\Psi (t,r,\phi,z)=0.
\label{3}
\end{equation}

In Ref. \cite{Silva}, author introduced an electromagnetic interactions \cite{WG,NB,TYW,ERFM} into the Klein-Gordon equation (\ref{2}) through the so called minimal substitution 
\begin{equation}
\partial_{\mu}\rightarrow D_{\mu}\equiv \partial_{\mu}+i\,e\,A_{\mu},
\label{4}
\end{equation}
where $e$ is the charge, and $A_{\mu}$ is the four-vector potential of electromagnetic field with
\begin{equation}
A_{\mu}=(A_0,-\vec{A})\quad,\quad \vec{A}=(0,A_{\phi},0).
\label{5}
\end{equation}
Here one can write the $0$-component of four-vector potential $e\,A_{0}=V$. In Ref. \cite{Silva}, author considered the electromagnetic four-vector potential given in the form (\ref{5}) assuming $(+,-,-,-)$ or $-2$ signature of the space-time geometry \cite{WG}. Worthwhile it is known in classical general relativity that the Som-Raychaudhuri space-time in the form (\ref{1}) and the Minkowski metric $ds^2_{Min}=-dt^2+r^2\,d\phi^2+dr^2+dz^2$ is of Lorentzian with $(-,+,+,+)$ or $+2$ signature. Therefore, the correct form of electromagnetic four-vector potential is $A_{\mu}=(-A_0,\vec{A})$ with its contravariant form $A^{\mu}=(A^0,\vec{A})$.

With this, that is, $\partial_{t}\rightarrow D_{t}$ and $\partial_{\phi}\rightarrow D_{\phi}$ using (\ref{4})-(\ref{5}), the KG-equation (\ref{3}) becomes
\begin{eqnarray}
&&[-(\partial_{t}+i\,e\,A_{0})^2+\frac{1}{r}\,\partial_{r}\,(r\,\partial_{r})+\{\frac{1}{r}\,(\partial_{\phi}-i\,e\,A_{\phi})-\Omega\,r\,(\partial_{t}+i\,e\,A_{0})\}^2\nonumber\\
&&+\partial_{z}^2-M^2]\,\Psi (t,r,\phi,z)=0.
\label{6}
\end{eqnarray}
The Aharonov-Bohm potential considered in Ref. \cite{Silva} (please see the paragraph in between Eqs. (3) and (4) in Ref. \cite{Silva}) is given by 
\begin{equation}
e\,A_{\phi}=-\frac{\Phi}{r}\quad,\quad \Phi=\frac{\Phi_B}{(2\,\pi/e)}, 
\label{7}
\end{equation}
where $\Phi_B$ is the Aharonov-Bohm magnetic flux. The above form of potential (\ref{7}) in literature  known as the Aharonov-Bohm-Coulomb (ABC) potential.

Therefore, the Eq. (\ref{6}) using (\ref{7}) can now be expressed as
\begin{eqnarray}
&&[-(\partial_{t}+i\,V)^2+\frac{1}{r}\,\partial_{r}\,(r\,\partial_{r})+\{\frac{1}{r}\,(\partial_{\phi}+i\,\frac{\Phi}{r})-\Omega\,r\,(\partial_{t}+i\,V)\}^2\nonumber\\
&&-M^2+\partial_{z}^2]\,\Psi (t,r,\phi,z)=0.
\label{8}
\end{eqnarray}
Note that the above Eq. (\ref{8}) is different from the derived one Eq. (4) in Ref. \cite{Silva}.
 
The solution to Eq. (\ref{8}) is as follow:
\begin{equation}
\Psi (t, r, \phi, z)=e^{i\,(-E\,t+l\,\phi)\pm k\,z}\,R (r),
\label{9}
\end{equation}
where $E$ is the energy of charged particle, $l=0,\pm 1, \pm 2,..$ are the eigenvalues of the $z$-component of the angular momentum operator, $k$ is a positive real number and the $\pm$ sign is chosen to coincide with the negative/positive $z$-axis to produce the decay factor $e^{-k\,|z|}$.

Substituting the solution (\ref{9}) into the Eq. (\ref{8}), we obtain
\begin{eqnarray}
&&R'' (r)+\frac{1}{r}\,R' (r)+[(E-V)^2-M^2+k^2-\frac{1}{r^2}\,\{(l+\frac{\Phi}{r})+\Omega\,r^2\,(E-V)\}^2]\,R (r)=0\nonumber\\\Rightarrow 
&&R'' (r) +\frac{1}{r}\,R' (r)+[(E-V)^2-M^2+k^2-\frac{(l+\frac{\Phi}{r})^2}{r^2}-\Omega^2\,r^2\,(E-V)^2\nonumber\\
&&-2\,\Omega\,(E-V)\,(l+\frac{\Phi}{r})]\,R (r)=0.
\label{10}
\end{eqnarray}
Substituting the Coulomb-type potential $V=\frac{\xi}{r}$ considered in Ref. \cite{Silva} where, $\xi=\pm\,e^2$ (please see the paragraph in between Eqs. (3) and (4) in Ref. \cite{Silva}) into the Eq. (\ref{10}), we obtain
\begin{equation}
[\frac{d^2}{dr^2}+\frac{1}{r}\,\frac{d}{dr}+\lambda-\frac{j^2}{r^2}-\frac{a}{r}+b\,r-\omega^2\,r^2-\frac{c}{r^3}-\frac{d}{r^4}]\,R (r)=0,
\label{11}
\end{equation}
where
\begin{eqnarray}
&&\lambda=E^2-\Omega^2\,\xi^2-M^2+k^2-2\,\Omega\,E\,l,\nonumber\\
&&j=\sqrt{l^2-\xi^2-2\,\Omega\,\xi\,\Phi},\nonumber\\
&&a=2\,E\,\xi+2\,\Omega\,(E\,\Phi-\xi\,l),\nonumber\\
&&b=2\,\Omega^2\,E\,\xi,\nonumber\\
&&\omega=\Omega\,E,\nonumber\\
&&c=2\,\Phi\,l,\nonumber\\
&&d=\Phi^2.
\label{12}
\end{eqnarray}
The above Eq. (\ref{11}) is completely different from Eq. (6) derived in Ref. \cite{Silva}. This fact jeopardize Eq. (6) derived in Ref. \cite{Silva} and therefore, the further analysis with this equation as well as the energy eigenvalues presented there are incorrect.

{\bf Conclusions:} Taking into account all the physical parameter considered in Ref. \cite{Silva}, we have derived the final form of the Klein-Gordon equation with the potential $e\,A_{\phi}=-\frac{\Phi}{r}$ (which is known as the Aharonov-Bohm-Coulomb potential) and Coulomb-type vector potential $V(r)=\frac{\xi}{r}$ in the background of Som-Raychaudhuri space-time. In this derived equation Eq.(\ref{11}), we have seen that there are two additional terms containing $\frac{1}{r^3}$ and $\frac{1}{r^4}$ which are missing in Eq. (6) derived in Ref. \cite{Silva}. The presence of the terms $\frac{1}{r^3}$ and $\frac{1}{r^4}$ makes the Eq. (\ref{11}) highly singular nature. To ease the solution, one may try to remove the $\frac{1}{r^3}$ singularity by selecting S-wave states ($l=0$). However, to remove both singularities $\frac{1}{r^3}$ and $\frac{1}{r^4}$ one has to choose $\Phi=0$, which changes the physics of the whole problem.

For $\Phi\rightarrow 0$, the Eq. (\ref{11}) redcues to
\begin{equation}
[\frac{d^2}{dr^2}+\frac{1}{r}\,\frac{d}{dr}+\lambda-\frac{j^2_{0}}{r^2}-\frac{a_0}{r}+b\,r-\omega^2\,r^2]\,R (r)=0,
\label{13}
\end{equation}
where
\begin{eqnarray}
&&\lambda=E^2-\Omega^2\,\xi^2-M^2+k^2-2\,\Omega\,E\,l,\nonumber\\
&&j_0=\sqrt{l^2-\xi^2},\nonumber\\
&&a_0=2\,\xi\,(E-\Omega\,l),\nonumber\\
&&b=2\,\Omega^2\,E\,\xi,\nonumber\\
&&\omega=\Omega\,E.
\label{14}
\end{eqnarray}
Equation (\ref{13}) is the radial wave-equation of Klein-Gordon charged particle in the Som-raychaudhuri space-time subject to only Coulomb-type vector potential. Thus, the Eq. (\ref{11}) derived here jeopardize Eq. (6) derived in Ref. \cite{Silva}, and the further analysis as well as the results obtained there are incorrect.

\section*{Acknowledgement}

Author acknowledge the anonymous kind referee(s) for their helpful and constructive suggestions.

\end{document}